\documentclass[prl,aps,showpacs,twocolumn,floatfix]{revtex4}
\usepackage{epsfig} \usepackage{graphics} \usepackage{bm}
\usepackage{amssymb}
\begin{document}

\title{Equivalence between Gravitational Mass and Energy for a Quantum Body at a Macroscopic Level}

\author{A.G. Lebed*}

\affiliation{Department of Physics, University of Arizona, 1118 E.
4-th Street, Tucson, AZ 85721, USA}
\date{\today}

\begin{abstract}
We define passive gravitational mass operator of the simplest composite quantum body - a hydrogen atom - to be proportional to its weight operator in a weak gravitational field. Although it does not commute with energy operator, taken in the absence of the field, the equivalence between the expectation value of passive gravitational mass and energy is shown to survive for stationary quantum states. All the so-called relativistic corrections to electron energy in a hydrogen atom are taken into account in the calculations.
\end{abstract}

\maketitle

\pagebreak

Formulation of a successful quantum gravitational theory is considered
to be one of the most important problems in physics and the major
step towards the so-called "Theory of Everything".
On the other hand, fundamentals of general relativity and quantum
mechanics are so different that it is possible that these two theories will not be united in the feasible future.
In this difficult situation, it seems to be important to suggest a
combination of quantum mechanics and some non-trivial approximation
of general relativity.
In particular, this is important in the case where such theory leads to
some meaningful physical results and, thus, deepens our understanding
of physics.

A notion of gravitational mass of a composite body is known to be
non-trivial in general relativity and related to the following
paradoxes.
If we consider a free photon with energy $E$ and apply to it the
so-called Tolman formula for gravitational mass \cite{Landau}, we will
obtain $m^g=2E/c^2$ (i.e., two times bigger value than the expected
one) \cite{Misner}.
If a photon is confined in a box with mirrors, then we have a composite
body at rest.
In this case, as shown in Ref. \cite{Misner}, we have to take into account
a negative contribution to $m^g$ from stress in the box walls to restore
the Einstein's equation, $m^g=E/c^2$.
It is important that the later equation is restored only after averaging
over time.
A role of the classical virial theorem in establishing of the equivalence
between averaged over time gravitational mass and energy is discussed
in detail in Refs. \cite{Nordtvedt,Carlip} for different types of classical
composite bodies.
In particular, for electrostatically bound two bodies, it is shown
that gravitational field is coupled to a combination $3K+2U$,
where $K$ is kinetic energy, $U$ is the Coulomb potential
energy.
Since the classical virial theorem states that the following time average
is equal to zero, $\bigl< 2K+U \bigl>_t = 0$, then we conclude that
averaged over time gravitational mass is proportional to the total amount
of energy \cite{Nordtvedt,Carlip},
 \begin{equation}
\bigl< m^g \bigl>_t = m_1 + m_2 + \bigl<3K + 2U \bigl>_t/c^2 = E/c^2,
\end{equation}
where $m_1$ and $m_2$ are bare masses of the above mentioned bodies.

The main goal of our Letter is to study a quantum problem about passive
gravitational mass of a composite body.
As the simplest example, we consider a hydrogen atom in a weak gravitational field, where we take into account not only classical kinetic and Coulomb potential energies of an electron in the atom, but also the so-called relativistic corrections \cite{Schwabl}.
We claim two main results in the Letter.
Our first result is that the equivalence between passive gravitational
mass and energy survives at a macroscopic level, if we take into account only non-relativistic kinetic and Coulomb potential energies couplings with an external gravitational field, due to the quantum virial theorem \cite{Park}. Our second result is that the above-mentioned equivalence survives even in the case where we take into account the relativistic corrections to an electron motion in the atom, due to more complex theorem than the quantum virial one.
We pay attention that both of the above-mentioned results are non-trivial.
Indeed, below we define passive gravitational mass operator of an electron, $\hat m^g_e$, to be proportional to its weight operator in a weak gravitational field.
It is important that this gravitational mass operator of an electron occurs not to commute with its energy operator, taken in the absence of the field. Therefore, from the first point of view, it seems that the equivalence between passive gravitational mass and energy is broken.
Nevertheless, using rather sophisticated mathematical theorems, we show that the expectation value of passive gravitational mass operator is $< \hat m^g_e > = E'_i / c^2$ for stationary quantum states in a hydrogen atom, where $E'_i$ is the total electron energy of i-th atomic energy level, which includes the relativistic corrections.

Let us use the so-called weak field approximation to describe spacetime far enough from some massive classical body \cite{Misner-2,Lebed-2},
\begin{eqnarray}
d s^2 = -\biggl(1 + 2 \frac{\phi}{c^2} \biggl)(cdt)^2
+ \biggl(1 - 2 \frac{\phi}{c^2} \biggl) (dx^2 +dy^2+dz^2 ),
\nonumber\\
\phi = - \frac{GM}{R} ,
\end{eqnarray}
where $G$ is the gravitational constant, $c$ is the velocity of
light, $M$ is a mass of the body, $R$ is a distance from it.
Then, in the local proper spacetime coordinates,
\begin{eqnarray}
&&x'=\biggl(1-\frac{\phi}{c^2} \biggl) x, \ y'= \biggl(1-\frac{\phi}{c^2} \biggl) y,
\nonumber\\
&&z'=\biggl(1-\frac{\phi}{c^2} \biggl) z , \ t'= \biggl(1+\frac{\phi}{c^2} \biggl) t,
\end{eqnarray}
the non-relativistic Schr\"{o}dinger equation for electron motion in a hydrogen atom can be approximately written in the following standard form:
\begin{eqnarray}
i \hbar \frac{\partial \Psi({\bf r'},t')}{\partial t'} = \hat H_0 (\hat {\bf p'},{\bf r'}) \Psi({\bf r'},t')  ,
\nonumber\\
 \ \hat H_0 (\hat {\bf p'},{\bf r'}) = m_e c^2 + \frac{\hat {\bf p'}^2}{2m_e}-\frac{e^2}{r'} ,
\end{eqnarray}
where $m_e$ and $e$ are the bare electron mass and charge, respectively; $r'$ is a distance between electron and proton, $\hat {\bf p'} = - i \hbar \partial /\partial {\bf r'}$ is electron momentum operator in the local proper coordinates. We stress that, in Eq.(4) and everywhere below, we disregard all tidal effects (i.e., we do not differentiate gravitational potential with respect to electron coordinates, ${\bf r}$ and ${\bf r'}$, corresponding to electron positions in the center of mass coordinate systems). It is possible to show that this means that we consider a hydrogen atom as a point-like body and disregard all tidal terms in electron Hamiltonian. Note that the above mentioned tidal effects are usually very small and of the order of $(r_B/R_0)|\phi/c^2|  \sim 10^{-17}|\phi/c^2|$ in the Earth's gravitational field, where $r_B$ is the so-called Bohr radius and $R_0$ is the Earth's radius.

Below, we treat the weak gravitational field (2) as a perturbation in an inertial coordinate system, corresponding to spacetime coordinates $(x,y,z,t)$ in Eq.(3) \cite{Nordtvedt,Carlip}, and calculate the corresponding Hamiltonian:
\begin{equation}
\hat H_0(\phi,\hat {\bf p},{\bf r}) = m_e c^2 + \frac{\hat {\bf p}^2}{2m_e}-\frac{e^2}{r} + m_e  \phi
+ \biggl( 3 \frac{\hat {\bf p}^2}{2 m_e}
-2\frac{e^2}{r} \biggl) \frac{\phi}{c^2}.
\end{equation}
It is convenient to rewrite the Hamiltonian (5) in the following form:
\begin{equation}
\hat H_0(\phi,\hat {\bf p},{\bf r}) = m_e c^2 + \frac{\hat {\bf p}^2}{2m_e}-\frac{e^2}{r} + \hat m^g_e \phi \ ,
\end{equation}
where we introduce passive gravitational mass operator of an electron:
\begin{equation}
\hat m^g_e  = m_e  + \biggl(\frac{\hat {\bf
p}^2}{2m_e}  -\frac{e^2}{r}\biggl)/ c^2
+ \biggl(2 \frac{\hat {\bf p}^2}{2m_e}-\frac{e^2}{r} \biggl)/ c^2 \ ,
\end{equation}
which is proportional to its weight operator in the weak gravitational field (2). Note that, in Eq.(7), the first term corresponds to the bare electron mass, $m_e$, the second term corresponds to the expected electron energy
contribution to the gravitational mass operator, whereas the third non-trivial term is the virial contribution to passive gravitational mass operator.
It is possible to make sure \cite{Lebed-3} that Eqs.(6),(7) can be obtained directly from the Dirac equation in a curved spacetime, corresponding to the weak centrosymmetric gravitational field (2) (see, for example, \cite{Fischbach}), if we disregard all tidal terms.

Here, we discuss some important consequence of Eqs.(6),(7).
It is crucial that the operator (7) does not commute with electron energy
operator, taken in the absence of the field.
Therefore, it is not clear from the beginning that the equivalence between
electron passive gravitational mass and its energy exists. To established the
equivalence at a macroscopic level, we consider a macroscopic ensemble of hydrogen atoms with each of them being in a stationary quantum state with a definite energy $E_i$.
Then, from Eq.(7), it follows that the expectation value of electron passive gravitational mass operator per atom is
\begin{equation}
<\hat m^g_e > = m_e + \frac{ E_i}{c^2}  + \biggl< 2 \frac{\hat
{\bf p}^2}{2m_e}-\frac{e^2}{r} \biggl> /c^2 = \biggl(m_e + \frac{E_i}{c^2}
\biggl) ,
\end{equation}
where the third term in Eq.(8) is zero in accordance with the quantum virial theorem \cite{Park}.
Therefore, we conclude that the equivalence between passive gravitational mass and energy survives at a macroscopic level for stationary quantum states, if we consider only pairings of non-relativistic kinetic and Coulomb potential energies with an external gravitational field.

Below, we study a more general case, where the so-called relativistic corrections to an electron motion in a hydrogen atom are taken into account.
As well known \cite{Schwabl}, there exist three relativistic correction terms, which have different physical meaning. The total Hamiltonian can be written as
\begin{equation}
\hat H (\hat {\bf p},{\bf r}) = \hat H_0 ( \hat {\bf p},{\bf r}) + \hat H_1 (\hat {\bf p},{\bf r}) ,
\end{equation}
where
\begin{equation}
\hat H_1 (\hat {\bf p},{\bf r}) = \alpha \hat {\bf p}^4 + \beta \delta^3 ({\bf r}) + \gamma \frac{\hat {\bf S} \cdot \hat {\bf L} }{r^3} ,
\end{equation}
with the parameters $\alpha$, $\beta$, and $\gamma$ being:
\begin{equation}
\alpha = - \frac{1}{8 m_e^3 c^2}, \ \beta=\frac{\pi e^2 \hbar^2}{2m_e^2c^2}, \ \gamma = \frac{e^2}{2 m_e^2 c^2},
\end{equation}
[Here, $\delta^3 ({\bf r}) = \delta (x) \delta (y) \delta (z)$ is a three dimensional Dirac delta-function.]
Note that the first contribution in Eq.(10) is called the kinetic term, the second one is the so-called Darwin term, and the third is the spin-orbital interaction, where $\hat {\bf L} = - i \hbar [{\bf r} \times \partial / \partial {\bf r}]$ is electron angular momentum operator.
In the presence of the weak gravitational field (2), the Schr\"{o}dinger equation for electron motion in the local proper spacetime coordinates (3) can be approximately written as
\begin{equation}
i \hbar \frac{\partial \Psi({\bf r'},t')}{\partial t'} = [\hat H_0(\hat {\bf p'},{\bf r'})+ \hat H_1(\hat {\bf p'},{\bf r'})] \Psi ({\bf r',t'}).
\end{equation}
[Note that, as discussed above, we disregard everywhere all tidal terms.]

By means of the coordinates transformation (3), the corresponding Hamiltonian in an inertial coordinate system (x,y,z,t) can be expressed as
\begin{eqnarray}
\hat H(\phi, \hat {\bf p},{\bf r})= [\hat H_0(\hat {\bf p}, {\bf r}) + \hat H_1 (\hat {\bf p},{\bf r})] \biggl(1 + \frac{\phi}{c^2} \biggl)
\nonumber\\
+ \biggl(2 \frac{\hat {\bf p}^2}{2 m_e}-\frac{e^2}{r} + 4 \alpha \hat {\bf p}^4 + 3 \beta \delta^3({\bf r}) + 3 \gamma \frac{\hat {\bf S} \cdot \hat {\bf L} }{r^3} \biggl) \frac{\phi}{c^2} .
\end{eqnarray}
For the Hamiltonian (13), passive gravitational mass operator of an electron can be written in more complicated than Eq.(7) form:
\begin{eqnarray}
&&\hat m^g_e = m_e + \biggl( \frac{\hat {\bf p}^2}{2m_e} - \frac{e^2}{r}
+ \alpha \hat {\bf p}^4 + \beta \delta^3 ({\bf r}) + \gamma \frac{\hat {\bf S} \cdot \hat {\bf L} }{r^3} \biggl)/c^2
\nonumber\\
&&+ \biggl(2 \frac{\hat {\bf p}^2}{2 m_e}-\frac{e^2}{r} + 4 \alpha \hat {\bf p}^4 + 3 \beta \delta^3 ({\bf r}) + 3 \gamma \frac{\hat {\bf S} \cdot \hat {\bf L} }{r^3} \biggl)/c^2 .
\end{eqnarray}
Let us consider again a macroscopic ensemble of hydrogen atoms with each of them being in a stationary quantum state with a definite energy $E'_i$, where $E'_i$ takes into account the relativistic corrections (10) to electron energy. In this case, the expectation value of the electron mass operator (14) per atom can be written as
\begin{eqnarray}
&&<\hat m^g_e > = m_e + \frac{ E'_i}{c^2}
\nonumber\\
&&+ \biggl<2 \frac{\hat {\bf p}^2}{2 m_e}-\frac{e^2}{r} + 4 \alpha \hat {\bf p}^4 + 3 \beta \delta^3 ({\bf r}) + 3 \gamma \frac{\hat {\bf S} \cdot \hat {\bf L} }{r^3} \biggl>/c^2 .
\end{eqnarray}

Below, we show that the expectation value of the third term in Eq.(15) is zero and, therefore, the Einstein's equation, related the expectation value of gravitational mass and energy, can be applied to stationary quantum states.
Here, we define the so-called virial operator \cite{Park},
\begin{equation}
\hat G = \frac{1}{2} (\hat {\bf p}{\bf r} +{\bf r} \hat {\bf p}) ,
\end{equation}
and write the standard equation of motion for its expectation value:
\begin{equation}
\frac{d}{dt} \biggl< \hat G \biggl> = \biggl< [\hat H_0(\hat {\bf p},{\bf r})+ H_1(\hat {\bf p},{\bf r}), \hat G] \biggl> ,
\end{equation}
where $[\hat A, \hat B]$, as usual, stands for a commutator of two operators, $\hat A$ and $\hat B$.
If we consider only a stationary quantum state with definite energy, $E'_i$, then the derivative $d<\hat G>/dt$ in Eq.(17) has to be zero and, thus,
\begin{equation}
\biggl< [\hat H_0(\hat {\bf p},{\bf r})+ H_1(\hat {\bf p},{\bf r}), \hat G] \biggl> = 0 ,
\end{equation}
where the Hamiltonians $\hat H_0(\hat {\bf p},{\bf r})$ and $\hat H_1(\hat {\bf p},{\bf r})$ are defined by Eqs.(4),(10).
By means of rather lengthy but straightforward calculations it is possible to show that
\begin{eqnarray}
&&\frac{[\hat H_0(\hat {\bf p},{\bf r}), \hat G]}{-i \hbar}=  2 \frac{\hat {\bf p}^2}{2m_e}-\frac{e^2}{r} , \ \ \frac{[\alpha \hat {\bf p}^4,\hat G]}{-i \hbar} = 4 \alpha \hat {\bf p}^4 ,
\nonumber\\
&&\frac{[\beta \delta^3 ({\bf r}), \hat G]}{-i \hbar} = 3 \beta \delta^3 ({\bf r}), \ \
\frac{1}{-i \hbar} \biggl[ \gamma \frac{\hat {\bf S} \cdot \hat {\bf L}}{r^3} ,\hat G \biggl] =
3 \gamma \frac{\hat {\bf S} \cdot \hat {\bf L}}{r^3} ,
\end{eqnarray}
where we take into account the following equality: $x_i \frac{d \delta(x_i)}{d x_i} = - \delta (x_i)$.
As directly follows from Eqs.(18),(19),
\begin{equation}
\biggl<2 \frac{\hat {\bf p}^2}{2 m_e}-\frac{e^2}{r} + 4 \alpha \hat {\bf p}^4 + 3 \beta \delta^3 ({\bf r}) + 3 \gamma \frac{\hat {\bf S} \cdot \hat {\bf L} }{r^3} \biggl> =0,
\end{equation}
and, therefore, Eq.(15) can be rewritten in the Einstein's form:
\begin{equation}
<\hat m^g_e > = m_e + \frac{E'_i}{c^2} .
\end{equation}

Note that Eq.(21) directly establishes the equivalence between the expectation value of electron passive gravitational mass and its energy in a hydrogen atom, including the relativistic corrections. We speculate that such equivalence exists also for more complicated quantum systems, including many-body systems with arbitrary interactions of particles. These reveal and establish the physical meaning of a coupling of a macroscopic quantum test body with a weak gravitational field. On the other hand, we pay attention that we have established the equivalence of passive gravitational mass and energy only at a macroscopic level, since we considered the expectation value of the electron mass. There exist claims \cite{Lebed-4, Lebed-5} that the corresponding equivalence may be broken at a microscopic level.
In addition, there has appeared a paper \cite{Singleton}, where it is shown that the equivalence principle between gravitational and inertial masses can be broken due to the Unruh and Hawking radiations.

We are thankful to N.N. Bagmet, V.A. Belinski, and V.E. Zakharov for useful discussions. This work was supported by the NSF under Grant DMR-1104512.

$^*$Also, at L.D. Landau Institute for Theoretical Physics, RAS,
2 Kosygina Street, 117334 Moscow, Russia.


\begin{references}
\bibitem{Lebed-2}
We pay attention that to calculate the quantum Hamiltonian in a linear with respect to a small parameter $\phi(R) / c^2$ approximation, we do not need to keep the terms of the order of $[\phi(R)/c^2]^2$ in the metric (2), in contrast to the classical problem about perihelion precession of mercury's orbit \cite{Misner-2}.
\end{references}
\end{document}